\documentclass[fleqn,usenatbib,useAMS]{mnras}
\usepackage{newtxtext,newtxmath}
\usepackage[T1]{fontenc}
\usepackage{caption}
\usepackage{threeparttable}
\DeclareRobustCommand{\VAN}[3]{#2}
\let\VANthebibliography\thebibliography
\def\thebibliography{\DeclareRobustCommand{\VAN}[3]{##3}\VANthebibliography}

\def\ASCA{\textit{ASCA}}
\def\XMM-Newton{\textit{XMM-Newton}}
\def\XMM{\textit{XMM}}
\def\Chandra{\textit{Chandra}}
\def\Suzaku{\textit{Suzaku}}
\def\Swift{\textit{Swift}}
\usepackage[dvipdfmx]{graphicx}	
\usepackage{amsmath}	

\title[Long Term X-Ray Spectral Variation of the Wolf-Rayet Binary WR 102-1]{Long Term X-Ray Spectral Variation of the Wolf-Rayet Binary WR 102-1 in the Galactic bulge: evidence for wind distortion in the binary}

\author[ T.\ Nagatsuka  et al.]{
Tomoki Nagatsuka,$^{1,2}$\thanks{E-mail: nagatsuka@ac.jaxa.jp}
Yasuharu Sugawara,$^{1}$
Ken Ebisawa,$^{1,2}$
\\
$^{1}$Institute of Space and Astronautical Science(ISAS), Japan Aerospace Exploration Agency(JAXA), 3-1-1 Yoshinodai, Chuo-ku, Sagamihara, Kanagawa, 252-5210, Japan\\
$^{2}$Department of Astronomy, Graduate School of Science, The university of Tokyo, 7-3-1 Hongo, Bunkyo-ku, Tokyo 113-0033, Japan\\ 
}



\date{Accepted XXX. Received YYY; in original form ZZZ}

\pubyear{2015}

\begin{document}
\label{firstpage}
\pagerange{\pageref{firstpage}--\pageref{lastpage}}
\maketitle

\begin{abstract}
WR~102-1 was detected by \Suzaku\/ as a conspicuous point source in the 6.7 keV intensity map of the central region of the Milky Way.
The source was suggested as a possible Wolf-Rayet binary based on its X-ray and infrared spectral characteristics.
The iron line emission is  expected to originate in the Wolf-Rayet star's dynamic  stellar-wind when colliding the companion's mild stellar wind.
Here, we report the result of a long-term X-ray monitoring of WR~102-1 since 1998 using archival data of \textit{ASCA}, \textit{XMM-Newton}, \textit{Chandra}, \textit{Suzaku}, and \textit{Swift}
to reveal variations of the iron K-emission line  and the circumstellar absorption. 
Consequently,
we have detected significant redshifts of the iron K-emission line from the \textit{XMM-Newton} observation in March 2003 and the \textit{Suzaku}  observation in September 2006.
Furthermore, when the red-shift was observed, which suggests that the Wolf-Rayet star was in front of the companion star, the circumstellar absorption values were smaller than other periods. These results appear contrary to the expectation if the Wolf-Rayet's stellar wind is spherically symmetric, but may be understood if  the Wolf-Rayet star's stellar wind is significantly distorted  due to the rapid  orbital motion  near the periastron.

\end{abstract}




\begin{table*}
    	\caption{\textit{XMM}, \textit{Chandra} and \textit{Suzaku} Observation log.}
	\begin{center}
	\label{tab:table2}
 	\begin{threeparttable}
	\begin{tabular}{|c|c|c|c|c|c|} \hline
     Satellite/Abbreviation & Detector  & Obs.ID & Start Time(UT) &Exposure time & off-axis angle\tnote{*} \\
                            &           &       &  & (ks)&(arcmin)\\\hline
     XMM1 & PN,MOS1,MOS2 & 0112970201 & 2000/09/23 22:37 & 12.3 &11.7\\
     Chandra1& ACIS-S & 1036 & 2000/10/27 11:27 & 35.0&10.5\\
     XMM2 & PN,MOS2 & 0112971501 & 2001/04/01 07:27 & 14.3&13.4\\
     Chandra2 & ACIS-I & 2271 & 2001/07/16 05:34 & 10.4&9.5\\
              & ACIS-I  & 2274 & 2001/07/16 08:43 & 10.4&5.7\\
              & ACIS-I  & 2285 & 2001/07/16 18:08 & 10.4&7.9\\
     XMM3 & PN,MOS1,MOS2  & 0144220101 & 2003/03/12 17:54 & 31.7&10.4\\
     Suzaku & XIS  & 501040010 & 2006/09/21 17:29 & 61.4& 7.4\\
             & XIS  & 501040020 & 2006/09/24 05:03 & 44.9&7.4\\
     XMM4 & MOS1,MOS2  & 0764190101 & 2016/03/10 08:51 & 32.5&13.1\\
     XMM5& PN,MOS1,MOS2  & 0802410101 & 2018/04/02 00:59 & 95.0&10.7\\
     \hline
    \end{tabular}
   \begin{tablenotes}
     \item[*] Angular distance of the source from the optical axis.
     \end{tablenotes}
 	\end{threeparttable}
\end{center}
\end{table*}

\begin{table}
	\centering
	\caption{\ASCA\//GIS Observation log.}
	\label{tab:table4}
	\begin{threeparttable}
	\begin{tabular}{|c|c|c|c|c|c|} \hline
    Obs.ID & Start Time(UT) & Exp.\  time & off-axis angle\tnote{*} \\
           &                &  (ks)    & (arcmin) \\ \hline
      60001100 & 1993/10/01 21:35 & 17.5& 17.5\\
     52006000 & 1994/09/22 03:46 & 58.4 & 12.7\\
      52006001 & 1994/09/24 02:07 & 21.5 & 12.7\\
      54001000 & 1996/09/19 15:55 & 11.2 & 19.0\\
      56001100 & 1998/09/07 15:01 & 10.0 & 21.1\\\hline
   \end{tabular}
    \begin{tablenotes}
     \item[*] Angular distance of the source from the optical axis.
     \end{tablenotes}
 	\end{threeparttable}
\end{table}

\begin{table}
	\centering
	\caption{\Swift\//XRT Observation log.}
	\label{tab:table5}
	\scalebox{0.9}[1.0]{
	\begin{threeparttable}
	\begin{tabular}{|c|c|c|c|c|} \hline
     Abbreviation   & Obs.ID & Start Time (UT) & Exp.\  time& off-axis angle\tnote{*}\\
                    &        &            &  (ks)& (arcmin)\\
 \hline
     Swift1 & 00030938001 & 2007/05/13 15:33 & 4.5 & 6.8\\
      & 00030938002 & 2007/05/16 23:59 & 1.4 & 7.7\\
      & 00030938003 & 2007/07/03 16:17 & 1.4 & 3.9\\
      & 00030938004 & 2007/08/08 03:25 & 2.0 & 7.6\\
     Swift2 & 00030938025 & 2019-05-13 04:14 & 2.2 & 7.6\\
      & 00030938026 & 2019-05-27 18:46 & 3.0 & 10.3\\
      & 00030938027 & 2019-06-10 18:50 & 3.3 & 9.0\\
      & 00030938028 & 2019-06-24 10:58 & 3.0 & 9.6\\
      & 00030938029 & 2019-07-08 09:46 & 2.8 & 7.3\\
      & 00030938031 & 2019-07-25 14:27 & 1.1 & 7.3\\
      & 00030938032 & 2019-07-26 08:12 & 1.2 & 9.3\\
      & 00030938033 & 2019-08-05 04:02 & 3.0 & 5.9 \\
      & 00030938034 & 2019-08-20 12:17 & 3.0 & 7.5\\
      & 00030938035 & 2019-09-02 02:46 & 3.0 & 5.7\\
     Swift3 & 00030938036 & 2020-02-04 13:13& 2.8 & 8.1\\
      & 00030938037 & 2020-03-04 16:49 & 2.7 & 9.9\\
      & 00030938038 & 2020-04-03 02:46 & 2.8 & 6.2\\
      & 00030938039 & 2020-05-04 10:49 & 2.7 & 11.7\\
      & 00030938040 & 2020-06-04 01:31 & 2.8 & 9.4\\
      & 00030938041 & 2020-07-06 03:19 & 2.8 & 6.3 \\
      & 00030938042 & 2020-08-04 03:25 & 2.7 & 7.5\\
      & 00030938043 & 2020-09-04 03:49 & 3.0 & 7.0\\
    \hline
   \end{tabular}
       \begin{tablenotes}
     \item[*] Angular distance of the source from the optical axis.
     \end{tablenotes}
 	\end{threeparttable}
   }
\end{table}

\section{Introduction}\label{sec:1}
Wolf-Rayet (WR) stars are luminous hot stars exhibiting broad emission lines, and considered to be  in the final stage of the stellar evolution of massive stars with $\gtrsim$25 $M_\odot$ \citep[e.g.,][]{Rosslowe2015}.
Most WR stars are  ejecting strong stellar winds \citep[e.g.,][]{Meynet2005}, 
and a  significant fraction of the WR stars are in binaries \citep{van2001,NeugentK}.
 O-type stars also often emit stellar winds and  are found in  binaries \citep{Sana2012}.
In those WR binaries with  stellar-wind emitting companions, 
wind-wind  collisions  are expected, which would exhibit intriguing astrophysical 
phenomena.

In the WR binaries with O- or B-type companions, 
shock waves are  generated when the  WR star's strong stellar wind   collides violently with  the companion star's milder stellar wind. This heats up the plasma hot enough to emit X-rays.  The X-ray luminosity and energy spectrum are highly dependent on the circumstellar   distance, mass-loss rates, and velocities of the stellar winds.Thus,  X-ray   spectral analysis can reveal   these physical parameter variations with the orbital phase \cite[e.g., ][]{2015PASJ...67..121S}. In addition, Doppler shift of the emission lines, if detected, would  reveal  dynamics of the  stellar winds.

A bright point object was discovered in the 6.7 keV intensity map of the central region of the Milky Way by \textit{Suzaku}, and the source was  suggested as a possible WR binary 
from its X-ray and infrared characteristics \citep{Hyodo2008}. The source was also detected by 
\textit{Chandra}\footnote{\url{https://cxc.cfa.harvard.edu/csc/}} and \textit{XMM-Newton}\footnote{\url{http://xmm-catalog.irap.omp.eu/}},  designated as CXOU J174645.2--281547/4XMM J174645.2--281548, and listed in the Wolf-Rayet catalog by \cite{Rate2020} as  WR~102-1\footnote{\url{http://pacrowther.staff.shef.ac.uk/WRcat/}}. 

Because WR~102-1 is located near the Galactic Center
at the Galactic coordinates  ($l$, $b$)=(0.70480, 0.13723), the source has been often serendipitously in the fields of view of the  pointing observations targeting the Galactic Center.
We have found 11 observations  in total by \textit{XMM-Newton}, Chandra, and \textit{Suzaku} from September 2000 to April 2018, for a total of 358 ksec exposure. Futhermore, WR~102-1 was in the field of view of five \textit{ASCA} observations carried out from October 1993 to September 1998, and   46 observations by \textit{Swift} from May 2007 to November 2020.
Using these datasets, we have studied long-term variations of   X-ray 
spectra and circumstellar environments of WR~102-1.

In Section 2, we present observations and data reduction.  Data analysis and results are given  in Section 3. Section 4 is the discussion, based on
our finding  of  intriguing variations of the iron K-line emission. 

\section{Observation and Data Reduction}\label{sec:section 2}

Table \ref{tab:table2} shows the observation log by \textit{Suzaku}, \textit{Chandra} and \textit{XMM-Newton} between September 2000 and April 2018. There were 11 observations of about 358~ksec  in total.  We also show angular distance of the
source from the telescope bore-sights for each observation.
In addition to the datasets used by  \cite{Hyodo2008}, this study includes two new observations of \textit{XMM-Newton} in March 2016 and April 2018 (XMM 4 and 5 in Table \ref{tab:table2}) with a total of  127~ksec.
In addition, we use \ASCA/GIS and \Swift/XRT archival data,
where WR102-1 were coincidentally in the fields of view.
The logs of these observations are shown in Table \ref{tab:table4} and \ref{tab:table5}.

Although the source is located in a crowded region  of  variable sources
near the Galactic Center, we found no significant
contamination from nearby bright sources in these observations. 
Thus, we simply extracted the source events from the region centering around the source position
in the following analysis.

\begin{figure*}
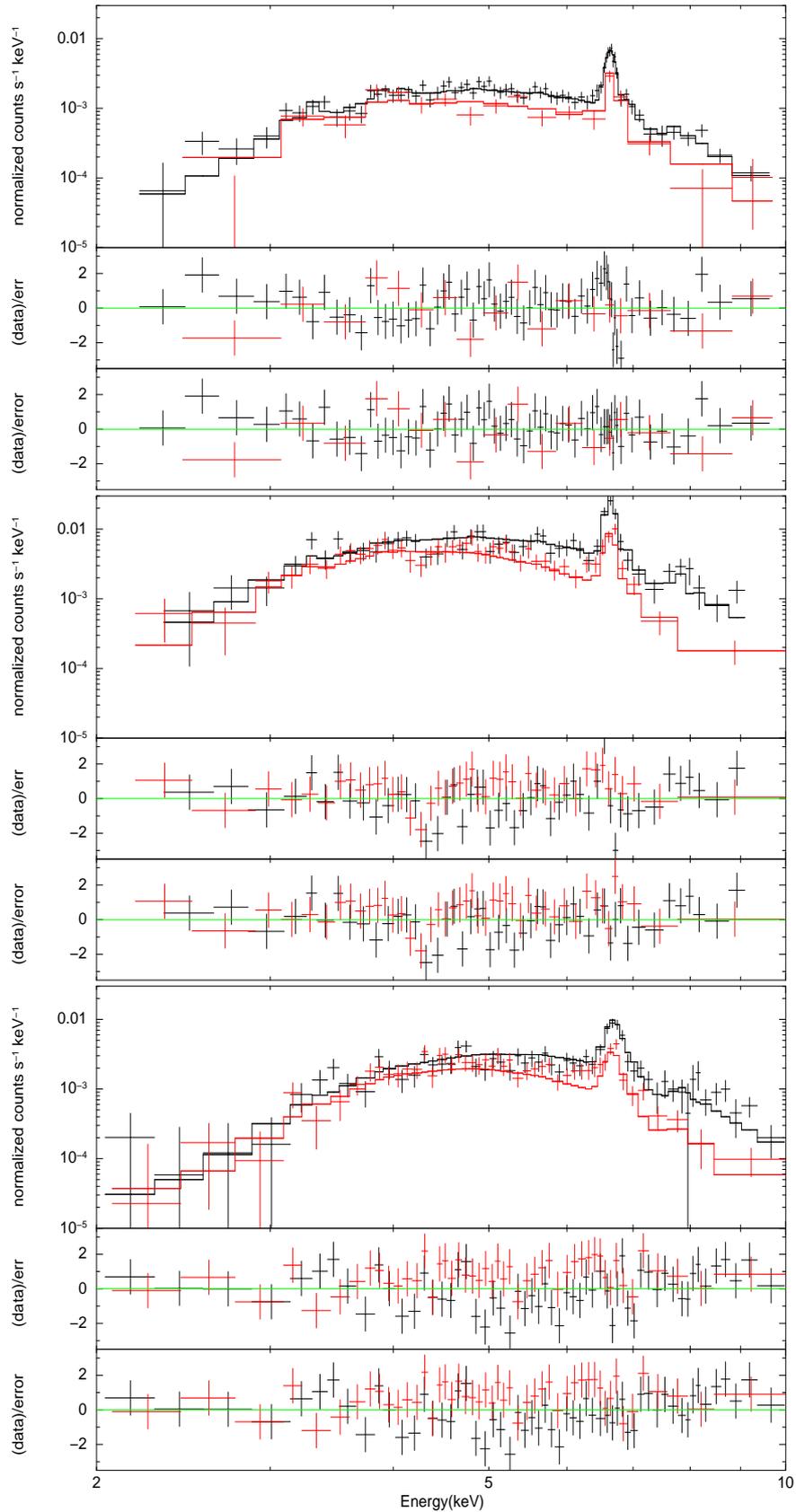

\begin{center}
    \includegraphics[width=0.4\textwidth, height=0.5\textheight, angle=270]{Suzaku}\\
    \includegraphics[width=0.4\textwidth,height=0.5\textheight,angle=270]{XMM3}\\
    \includegraphics[width=0.4\textwidth, height=0.5\textheight, angle=270]{XMM5_mod}\\
    \vspace*{1cm}
    \caption{Spectral fitting results  of WR~102-1 for Suzaku (top; black for FI, red for BI),  XMM3 (middle; black for PN, red for MOS1+2),  and XMM5 (bottom; same colors as for XMM3).
    In each of the three figures,  the upper-panel gives the observed spectra and best-fit models, the middle-panel gives the residual when the red-shift and the turbulent velocity are null, and the bottom-panel is the residual with the best-fit red-shifts and turbulent velocities.
    }
    \label{fig:spec}
    \end{center}
\end{figure*}

\begin{figure}
    \includegraphics[width=\columnwidth]{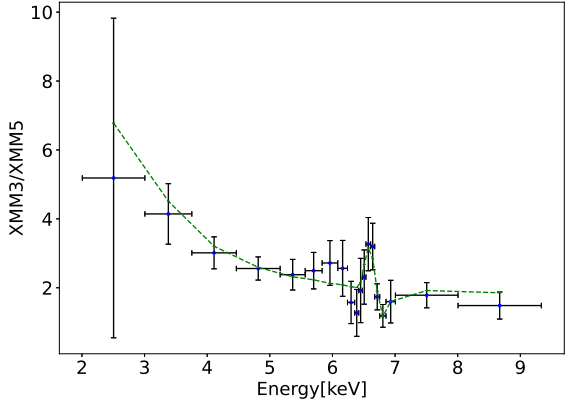}
    \caption{Spectral ratio of XMM3 to XMM5 (black). The green line shows the ratio of the best-fit model spectra  in Table 2. \label{fig:ratio}}
\end{figure}

\begin{figure}
     \includegraphics[width=\columnwidth]{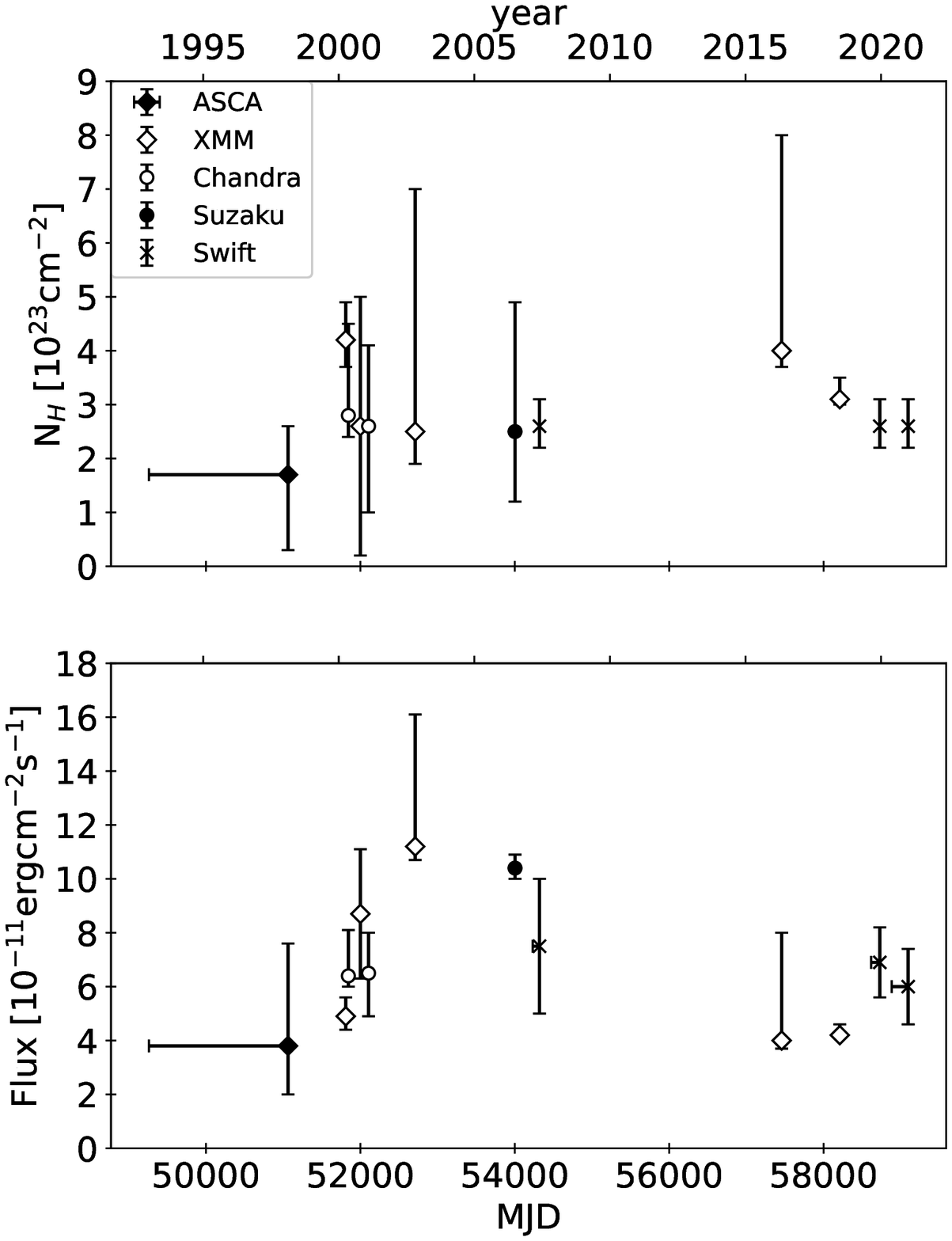}
     \caption{Variation of the X-ray flux (upper-panel) and the absorption  hydrogen column density   (lower-panel) by \textit{ASCA}, \textit{Suzaku}, \textit{XMM}, \textit{Chandra} and \textit{Swift}/XRT. \label{fig:lightcurve}
     }
\end{figure}

\subsection{\Suzaku}
 
X-ray Imaging Spectrometer \citep[XIS;][]{Koyama2007} on-board \textit{Suzaku} consists of four X-ray CCD cameras (XIS-0, XIS-1, XIS-2, XIS-3).  Together with the X-ray telescope \citep[][]{Serlemitsos2007}, the energy range is 0.2--12~keV. XIS-0, 2 and 3 hold front-illuminated (FI) CCDs and only XIS-1 carries a back-illuminated (BI) CCD.
The XIS data were reduced with HEASOFT version 6.26 \citep{2014ascl.soft08004N}, following the \Suzaku\/ data analysis  guide\footnote{\url{https://heasarc.gsfc.nasa.gov/docs/suzaku/analysis/abc/}}.
We used {\tt xselect} to extract the data products from the  
cleaned event files where background events had been already removed.
The source extraction region was a 20 arcsec radius circle, and the background was taken from a 60 arcsec circle in a nearby blank field.
The response functions  were calculated by \texttt{xisrmfgen} and \texttt{xisarfgen} \citep{Ishisaki2007}.
Different FI CCD data  were combined because they have almost identical performances.

\subsection{\Chandra}
The \textit{Chandra} satellite \citep{Weisskopf2001} is equipped with an X-ray telescope with a high angular resolution of 0.5 arcsecond. We used Advanced CCD Imaging Spectrometer (ACIS), which  consists of 10 CCDs; 2 × 2 square array of ACIS-I and 1 × 6 serial array of ACIS-S.
Two of the ACIS-S CCDs  are BI and the other eight are FI.
WR~102-1 was observed by ACIS-S on 27 October 2000 ("Chandra1" in Table \ref{tab:table2}), and by ACIS-I on 16 July 2001 ("Chandra2"). All the ACIS observations were conducted  in the Faint mode.

Chandra data was analyzed with CIAO version 4.11 \citep{2006SPIE.6270E..1VF} using the Level 1 mask file with PHA filtering. The source light curves were checked to confirm that there were no background flares.
We  used {\tt specextract} to extract the spectra. 
 The source region was $\sim$5 arcsec radius and the 
 background area was $\sim$15 arcsec radius in a nearby blank field.

\subsection{\XMM-Newton}
For \textit{XMM-Newton} \citep{Jansen2001}, we used EPIC, 
which is composed of two  front-illuminated CCD cameras (MOS1 and MOS2) and a back-illuminated  camera (PN).
Among the five XMM observations in Table \ref{tab:table2}, 
 WR~102-1 was out of the field of view of MOS1 in the observation of 1 April 2001 ("XMM2") and out of the field of view of PN in the observation of 10 March 2016 ("XMM4").  All the observations were carried out with the Medium filter.
The data were reduced with SAS version 17.0.0 \citep{2004ASPC..314..759G}, and  {\tt evselect} was used
to obtain the filtered event files for EPIC-MOS1, 2 and PN in 0.3–10.0 keV.
Light-curves were created  with {\tt evselect} and  the GTI file
was made with {\tt tabgtigen}.  Background flares were checked and removed.
The source area was  $\sim$30 arcsec radius and the background area was $\sim$90 arcsec radius in a nearby blank field.  MOS1 and  MOS2 spectra
and responces 
were combined with {\tt epicspeccombine}, and PN and MOS spectra were fitted simultaneously.

\subsection{\ASCA}
The \ASCA\/ observation log
is  shown  in  Tables \ref{tab:table4}.
Since the source flux was near the detection limit, we merged all the five observations carried out between October 1993 and September 1998, 
and  combined the GIS2 and GIS3 counters.
We used HEASOFT version 6.26, and {\tt xselect} to extract the data products from the  
cleaned event files where the background events had been already removed.
The source area was taken as $\sim$140 arcsec radius and the background was $\sim$400 arcsec radius in a nearby blank field.

\subsection{\Swift}

 The \Swift\/  observation log is  shown  in  Table \ref{tab:table5}.
We selected the \Swift\/-XRT data whose center positions are within 15 arcmin from WR102-1 and the exposure times  exceeding  1,000s.
We  produced the \Swift\/-XRT light-curves and spectra using the \textit{Swift}–XRT data product generator \citep{Evans2007,Evans2009}.
\footnote{\url{https://www.swift.ac.uk/user_objects/}}
We combined the snap-shot observations carried out in 2007, 2019, and 
2020, and created the three datasets  (the first column in Table \ref{tab:table5}).

\section{Data analysis and Results}\label{sec:section 3}

We created the  energy spectra and responses of \textit{XMM}, \textit{Chandra}, \textit{Suzaku}, and \Swift\/XRT according to the first column of Table 1 and \ref{tab:table5}.
These energy spectra were binned at a minimum of 15 counts per new bin for spectral fitting.

The spectral fitting was performed with {\tt xspec} package   
 \citep{1996ASPC..101...17A} using the {\tt tbabs} absorption model by  \cite{2000ApJ...542..914W} and optically thin thermal plasma model {\tt bvapec}  \citep{Smith2001}. 
We adopted the  Chi-Squared statistics  and the solar abundances by 
\citet{1989GeCoA..53..197A}.
The heavy element abundance ratios were fixed to the ones determined previously by  \textit{Suzaku} and \textit{XMM} analysis \citep[Table 2 in][]{Hyodo2008}. 

The three spectra, Suzaku, XMM3 and XMM5 have long enough exposures  and high photon statistics to enable detailed spectral analysis.  For these spectra, we carried out  model fitting allowing all the parameters free, including the plasma red-shifts and turbulent velocities.
We show  the spectral fitting results
for these spectra in Figure \ref{fig:spec}. 
In each spectrum, the top panel shows the spectra and the best-fit models.
The middle panel is the residual of the model fit with the  red-shifts and
 turbulent velocities being null, where
 we found   residuals at around  the iron emission line. If we allow the
 plasma red-shifts and turbulent velocities to be free, the residuals disappear (bottom panels in Figure \ref{fig:spec}).
 For Suzaku and XMM3, significant red-shifts are recognized, and there is a hint of blue-shift in XMM5.
 Best-fit parameters are shown in Table 2.

Figure \ref{fig:ratio} shows the spectral ratio of XMM3, which shows a clear energy red-shift,  to XMM5, which shows a hint of blue-shift. From this figure, changes of the line centroid energy is clearly recognized. 
Furthermore, this figure, as well as Table 2, indicates that the low-energy absorption is less significant in XMM3, when the iron emission line is red-shifted.  Possible origin of this is discussed in the next section.

For other spectra, rather low photon statistics does not allow to constrain the
red-shift and turbulent velocities, thus we made these parameters null.
In addition, for the spectra with low photon statistics 
in which the hydrogen column density or  temperature were not constrained (XMM2, Swift 1, 2 and 3), they were fixed to the average values of Suzaku, XMM3 and XMM5.

 Table \ref{tab:table3} lists  the best-fit parameter values and 90\% confidence intervals. XMM2 and Swift1, 2, and 3 were fixed at the average of the best-fit values of Suzaku, XMM3, and XMM5, which have high photon statistics. 
Figure \ref{fig:lightcurve} shows the long-term variations of the absorbed flux and hydrogen column densities. We see the source has been visible for the entire period from early 1990's to 2020.  The absorbed flux is variable by a factor of 2 to 3, and the column density is also significantly variable.

\begin{table*}
{
	\begin{center}
	\caption{Best-fit parameters of the spectral fitting.}
	\label{tab:table3}
    \begin{threeparttable}
	\begin{tabular}{|l|c|c|c|c|c|c|c|} \hline
  Observation & ${N}_{\rm H}$ & k$T$ & Red Shift & Velocity & Flux@2.0-8.0keV & $\chi ^2$/dof \\ & [$10^{23}$ cm$^{-2}$] & [keV] & $10^{-3}$ & [km~s$^{-1}$] & [$10^{-13}$ ergs~cm$^{-2}$~s$^{-1}$] \\\hline
     ASCA & 1.7$^{+2.5}_{-0.9}$ & 1.1$^{+1.4}_{-0.7}$& -- & -- &  3.8$^{+1.8}_{-3.8}$ & 161.5/101 \vspace{1mm}\\
     XMM1 & 4.2$^{+0.8}_{-0.6}$ & 6.0$^{+1.7}_{-1.5}$ & -- & -- & 4.9$^{+0.5}_{-0.7}$ & 53.48/31  \vspace{1mm}\\
     Chandra1 & 2.8$\pm$0.4 & 6.5$^{+6.1}_{-2.5}$ & -- & -- & 6.4 $^{+0.4}_{-1.7}$ & 14.6/13 \vspace{1mm}\\
     XMM2 & (2.6)\tnote{*} & (4.3)\tnote{*} & - & - & 8.7$^{+2.4}_{-2.4}$ & 0.69/3 \vspace{1mm}\\
     Chandra2 & 2.6$\pm$0.5 & 5.9$^{+3.4}_{-2.4}$ & -- & -- & 6.5$^{+1.6}_{-1.5}$ & 11.2/14 \vspace{1mm}\\
     XMM3 & 2.5$^{+0.6}_{-0.4}$  & 3.6$^{+1.6}_{-1.4}$ & 5.3$\pm^{5.8}_{5.3}$ & 3300$\pm$2700 & 11.2$^{+0.5}_{-4.9}$ & 46.5/57 \vspace{1mm}\\
     Suzaku & 2.5$^{+0.6}_{-0.4}$ & 3.6$^{+1.6}_{-1.4}$ & 5.1$\pm$1.5 & 300$^{+1700}_{-300}$& 10.4$^{+0.4}_{-0.5}$ & 64.7/82 \vspace{1mm}\\
     Swift1 & (2.6)\tnote{*} & (4.3)\tnote{*} & -- & -- & 7.5$\pm$2.5 & 0.49/1 \vspace{1mm}\\
     XMM4 & 4.0$^{+1.4}_{-1.1}$ & 4.8$^{+3.1}_{-2.2}$ & - & - &4.0$^{+0.3}_{-4.0}$ & 16.5/13 \vspace{1mm}\\
     XMM5 & 3.1$^{+0.2}_{-0.3}$ & 4.9$\pm$0.8 & $-$0.1$^{+3.3}_{-3.1}$  & 3200$\pm$1200 & 4.2$^{+0.1}_{-0.4}$ & 126.8/102 \vspace{1mm}\\
     Swift2 & (2.6)\tnote{*} & (4.3)\tnote{*} & -- & -- & 6.9$\pm$1.3& 6.87/6 \vspace{1mm}\\
     Swift3 & (2.6)\tnote{*} & (4.3)\tnote{*} & -- & -- & 6.0$\pm$1.4 & 5.25/4 \vspace{1mm}\\
     \hline
   \end{tabular}
\begin{tablenotes}
     \item[*] Those values in the parentheses are fixed at the average of the best-fit values of Suzaku, XMM3, and XMM5.
\end{tablenotes}
     \end{threeparttable}
   \end{center}
}
\end{table*}

\section{Discussion}\label{sec:section 4}
 We have carried out a  long-term X-ray spectral study of WR~102-1 using \ASCA, \Suzaku, \Chandra, \XMM\/ and \Swift/XRT.  As a result,  we have confirmed  significant variation of the X-ray flux, low-energy absorption, and, in particular, iron line profile over two decades. In this section we discuss  interpretation of these variations based on simple assumptions.
 
First assumption is that an increase of the soft X-ray absorption suggests  an increase of the WR stellar wind density in the line-of-sight. Next, the X-ray luminosity due to the stellar wind-wind collision  is expected to vary along  with the binary separation acoording to
the  inversely proportional relation \citep{Usov1992}. Thus, dimming  the X-ray radiation suggests  increase of  the binary separation.
Finally, 
  X-rays are considered to be emitted from the cone-shape bow shock, which takes place when the  WR star's strong stellar wind collides the O-star's milder stellar wind, so that the flow direction at the bow shock is from the WR star to the companion star. Motion of the X-ray emitting plasma flow at the bow shock  results in the spectral line broadening and the energy shift.

From these interpretations, the binary geometries at Suzaku, XMM3 and XMM5 observations may be  as shown in Figure 4.
In Suzaku and XMM3 (upper-panel), the binary separation is smaller thus the X-ray luminosity is higher. The WR-star is in front of the companion star and the plasma flow at the bow shock is receding, thus the iron line red-shifts are  seen. The WR star's wind is significantly distorted (see below).
In XMM5 (lower-panel),  the binary separation is larger, and the WR stellar wind is closer to spherically symmetric, as it is not significantly affected by the binary motion.  X-rays are more significantly absorbed by the WR star's  wind.

It is known that the binary  stellar wind is significantly distorted near the periastron because of the proximity of the stars and  the rapid orbital motion  \citep[e.g.,][]{2008MNRAS.388L..39O}. In particular,    a "hole" may be produced in the WR star's stellar wind.
If we are observing the X-rays from the bow shock through the hole in the stellar wind (Figure 4 top), the soft X-ray absorption will be less significant.

\cite{Hyodo2008} suggested that WR~102-1 may be a WC$+$O colliding wind binary with an elliptical orbit. Our results not only support their models, but also suggest that WR~102-1 is a long-period binary  with a period of at least 10 years. We believe that continuous X-ray monitors over the next years will help prove this hypothesis.

\begin{figure}
    \includegraphics[width=\columnwidth]{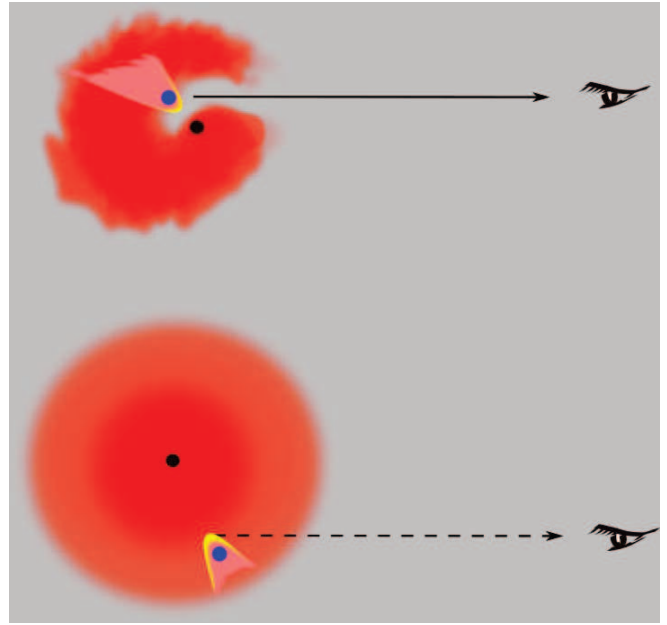}
    \caption{A possible binary geometry of WR~102-1. The black point is the WR star, the blue point is  the companion star, and the red region denotes the  stellar wind from the WR star. The X-rays are emitted from the cone-shape bow shock (yellow), which takes place when the  WR star's strong stellar wind collides the milder stellar wind from the O-star, so that the flow at the bow shock is from the WR star to the companion star.  Upper panel is for Suzaku and XMM3, when the clear red-shifts are  seen, the fluxes are  higher and the soft X-ray absorptions are  less, and the bottom-panel is for XMM5, when there is a hint of blue-shift, the flux is lower, and the X-rays are more absorbed. }
\end{figure}

\section{DATA AVAILABILITY}

The data analyzed in this research are all available at  HEASARC  of NASA\footnote{\url{https://heasarc.gsfc.nasa.gov}} and DARTS of ISAS/JAXA.\footnote{\url{https://darts.isas.jaxa.jp}} 

\section{Acknowledgements}

This research  has made use of data, software and web tools obtained from the High Energy Astrophysics Science Archive Research Center (HEASARC), a service of the Astrophysics Science Division at NASA/GSFC and of the Smithsonian Astrophysical Observatory's High Energy Astrophysics Division. This research also made use of data obtained from Data ARchives and Transmission System (DARTS), provided by Center for Science-satellite Operation and Data Archive (C-SODA) at ISAS/JAXA.This research made use of data supplied by the UK Swift Science Data Centre at the University of Leicester. We acknowledge the use of public data from the Swift data archive. Based on observations obtained with \textit{XMM-Newton}, an ESA science mission with instruments and contributions directly funded by ESA Member States and NASA.  We thank Ms.\ Juriko Ebisawa for the artwork of Figure 4.






\bibliographystyle{mnras}
\bibliography{WolfRayet} 








\bsp	
\label{lastpage}
\end{document}